\documentclass[
 reprint,
 superscriptaddress,
 nofootinbib,
 amsmath,amssymb,
 aps,
 prl,
 floatfix
]{revtex4-2}

\usepackage{braket}
\usepackage{amsmath}
\usepackage{graphicx}
\usepackage{dcolumn}
\usepackage{bm}
\usepackage{hyperref}
\hypersetup{
    colorlinks=true,
    linkcolor=blue,
    citecolor=blue,
    urlcolor=blue
}
\AtBeginDocument{%
}
\makeatletter
\AtBeginDocument{%
  \renewcommand{\equationautorefname}{Eq.\@autoref@insert@tagform}%
  \def\@autoref@insert@tagform~#1\null{~(#1)\null}%
}
\makeatother
\AtBeginDocument{%
}
\usepackage{siunitx}
\usepackage{physics}
\usepackage[table]{xcolor}
\usepackage{setspace}
\usepackage{booktabs}
\usepackage{color,soul}
\usepackage{diagbox}
\usepackage{isotope}
\usepackage{orcidlink}
\usepackage{comment}
\usepackage{pgfplots}
\pgfplotsset{compat=1.18}
\usepgfplotslibrary{colorbrewer} 
\pgfplotsset{
    colormap name=viridis,
}
\usepgfplotslibrary{groupplots}
\usepackage{tikz}
\usetikzlibrary{intersections}
\usetikzlibrary{decorations.pathmorphing}

\def\equationautorefname{Eq.}

\newcommand{\abinitio}{\textit{ab initio}~}

\newcommand{\NNb}{$N\bar{N}$~}

\allowdisplaybreaks

\begin{document}

\title{\texorpdfstring{\textit{Ab initio} evidence for surface-dominated antiproton annihilation in ${}^4 \rm He$} {\textit{Ab initio} evidence for surface-dominated antiproton annihilation in ${}^4 \rm He$} }

\author{Alireza Dehghani\orcidlink{0000-0003-2350-1433}}
\email{Contact author: alireza.dehghani@ijclab.in2p3.fr}
\affiliation{Universit\'e Paris-Saclay, CNRS/IN2P3, IJCLab, 91405 Orsay, France}%
\author{Guillaume Hupin\orcidlink{0000-0002-4285-7411}}
\affiliation{Universit\'e Paris-Saclay, CNRS/IN2P3, IJCLab, 91405 Orsay, France}%
\author{Sofia Quaglioni\orcidlink{0000-0002-7512-605X}}
\affiliation{Lawrence Livermore National Laboratory, P.O. Box 808, L-414, Livermore, California 94551, USA}%
\author{Petr Navr\'atil\orcidlink{0000-0002-7493-5293}}
\affiliation{TRIUMF, 4004 Wesbrook Mall, Vancouver, British Columbia, V6T 2A3, Canada}%

\date{\today}
\begin{abstract}
Low-energy antiproton beams at CERN/AD open the possibility of probing exotic nuclear structure through annihilation at the nuclear surface. 
The use of antiprotons as the probe for the nuclear surface is based on the assumption that the annihilation takes place at the periphery of the target. 
We test this idea for the lightest tightly bound nucleus, i.e., ${}^4 \mathrm{He}$, using the \abinitio no-core shell model combined with the resonating group method (NCSM/RGM), adapted to antiproton-nucleus dynamics.
After validating our microscopic calculations against available atomic and scattering data, we use the microscopic annihilation density to examine where annihilation occurs inside the antiprotonic atom. We find that the annihilation peaks in the tail of the ${}^{4}\mathrm{He}$ density, around $r\approx 2$ fm, and is strongly suppressed in the nuclear interior. Although this density is representation dependent, a similarity renormalization group (SRG) analysis of the NCSM/RGM Hamiltonian shows that the low-energy annihilation strength remains localized at large intercluster distances. These results support the phenomenological picture underlying antiprotonic-atom experiments: in a tightly bound system such as ${}^{4}\mathrm{He}$, antiproton annihilation is predominantly peripheral and is therefore sensitive to the nuclear-density tail.
\end{abstract}

\maketitle

The availability of low-energy antiproton beams at the CERN antiproton decelerator (CERN/AD), together with new experimental proposals at CERN~\cite{caravita2025cern}, has renewed interest in the theoretical study of antiprotonic systems~\cite{Duerinck:2026otx,duerinck2023antiproton, Vorabbi:2019ciy}. One experiment of particular relevance to low-energy nuclear physics is the antiProton Unstable Matter Annihilation (PUMA)~\cite{puma}, which aims to investigate the surface properties of stable and rare isotopes using low-energy antiprotons. The experiment is built around the idea that antiprotonic probes offer a unique sensitivity to the tail of the nuclear density, making them particularly well-suited to the study of surface phenomena such as halo structures and neutron skins~\cite{puma, Trzcinska:2001sy, Lubinski:1998xf}. \par
However, a reliable interpretation of the experimental data requires a detailed knowledge of the annihilation process, from antiproton capture and atomic cascade down to the lowest atomic orbits, to the subsequent non-perturbative coupling between the antiprotonic atomic state and the nucleus, ultimately leading to annihilation. The latter stage is the focus of the present work. Specifically, we aim to clarify where the annihilation process is spatially localized and whether it predominantly occurs at the nuclear periphery.\par 
To answer this question, we present an \abinitio reaction calculation of antiprotonic ${}^{4}\mathrm{He}$. Our goal is to test, in the lightest tightly bound nucleus, whether low-energy antiproton annihilation is a peripheral process governed by the tail of the nuclear density, thereby testing the central assumption underlying antiprotonic-atom probes such as PUMA. After ascertaining that the adopted microscopic Hamiltonian gives a consistent description of the available $\bar{p}$-${}^4 \rm He$ data, we calculate the annihilation density, which characterizes the antiproton annihilation strength as a function of target-antiproton separation. We observe that the annihilation density is suppressed at the nuclear interior and reaches its maximum at the tail of the nuclear density.\par
To study the $\bar{p}$-${}^4 \mathrm{He}$ system, we describe the ${}^4 \mathrm{He}$ target using the translationally invariant no-core shell model (NCSM)~\cite{Barrett:2013nh, Navr_til_2000,navratil2008ab}, and the $\bar{p}$-${}^4 \mathrm{He}$ dynamics using the no-core shell model/resonating group method (NCSM/RGM)~\cite{Quaglioni:2008sm,quaglioni, unified}. The NCSM/RGM formalism and its extension to antiproton-nucleus systems are presented in detail in Refs.~\cite{Quaglioni:2008sm,paperprc}. In this method, the ansatz for the total wave function of the system can be written as
\begin{equation}
    \psi^{J^\pi T}=\sum_{\nu} \int dr r^2  \frac{u^{J^\pi T}_\nu(r)}{r}  \left| \phi_{\nu r}^{J^\pi T}\right\rangle,
\label{eq:rgm_ansatz_integral_form}
\end{equation}
where $u^{J^\pi T}_\nu(r)$ is the radial part of the relative motion amplitude and $| \phi_{\nu r}^{J^\pi T}\rangle$ contains the eigenstates of the target and projectile together with the angular dependence of the relative motion. The collective index $\nu$ denotes the channel index and comprises the quantum numbers of the target, projectile, and the relative motion. 
The unknown relative motion amplitudes in \autoref{eq:rgm_ansatz_integral_form} satisfy the RGM's integrodifferential equation, which includes nonlocal strong and Coulomb potential kernels. The strong-interaction contribution can be written as
\begin{align}
    V^s_{\nu'\nu}(r',r)&=\left\langle \phi_{\nu' r'}^{J^\pi T}\middle| V_s \middle| \phi_{\nu r}^{J^\pi T}\right \rangle \notag\\& = (A-1)\sum_{nn'}^{N_{\text{max}}} R_{n\ell,b}(r) R_{n'\ell',b}(r') \notag\\& \quad \times  \left\langle \phi_{\nu' n',b}^{J^\pi T}\middle| V_{A-1,A} \middle| \phi_{\nu n,b}^{J^\pi T}\right \rangle,
\label{eq:direct_def}
\end{align}
Here $A$ is the total number of particles, $R_{n\ell,b}(r)$ denotes a harmonic oscillator (HO) radial function with oscillator length $b$, $n$ is the radial HO quantum number, $V_{A-1,A}$ is the strong potential between an antiproton and a target nucleon, and $| \phi_{\nu n,b}^{J^\pi T} \rangle$ is the NCSM/RGM basis state. The truncation parameter $N_{\text{max}}$ in \autoref{eq:direct_def} is chosen consistently with the many-body truncation $N^{\text{cluster}}_{\text{max}}$ used for the NCSM wave function of the target. \par
We solve the RGM's integrodifferential equation for both bound and scattering states using the calculable $R$-matrix method for complex-valued potentials~\cite{descouvemont2010r,hesse1998coupled}. Unless stated otherwise, we use a channel radius of $a_c=30$ fm and a Lagrange mesh with $n_s=100$ points. For antiprotonic-atom wave functions, a much larger channel radius is required to accommodate their spatial extent. Furthermore, we use the HO frequency $\hbar \omega=20$ MeV for both the NCSM and NCSM/RGM calculations. As in our earlier work, we suppress the long-range numerical artifacts in the NCSM/RGM potential kernels (e.g., the strong kernel in \autoref{eq:direct_def}) by introducing multiplicative regulators applied directly to the strong and Coulomb kernels, and denoting the corresponding cutoff radii by $r_{\text{reg}}$ and $r_{\text{reg,c}}$, respectively. \par
The target wave functions are obtained using an antisymmetrized HO basis truncated at $N^{\text{cluster}}_{\text{max}}$. As input, we use the two-body N${}^3$LO $\chi$EFT potential of Ref.~\cite{entem2003accurate}. In the current application, we neglect chiral three-nucleon forces (3NFs). While their inclusion may affect quantitative details, we expect the main qualitative conclusions, such as the peripheral character of annihilation and the overall scale of the strong-interaction level shifts, to remain unchanged. For the $N\bar{N}$ interaction, we choose the Kohno-Weise (KW) optical potential model~\cite{kohno1986proton}. The long-range part of this potential is derived from the Ueda meson-exchange nucleon-nucleon ($NN$) interaction~\cite{ueda1979antinucleon} by means of $G$-parity transformation~\cite{richard2020antiproton}. At short distances, a phenomenological imaginary core of Woods-Saxon form is used to simulate matter-antimatter annihilation. Other choices for the $N\bar{N}$ interaction are discussed in Ref.~\cite{carbonell2023comparison}. \par
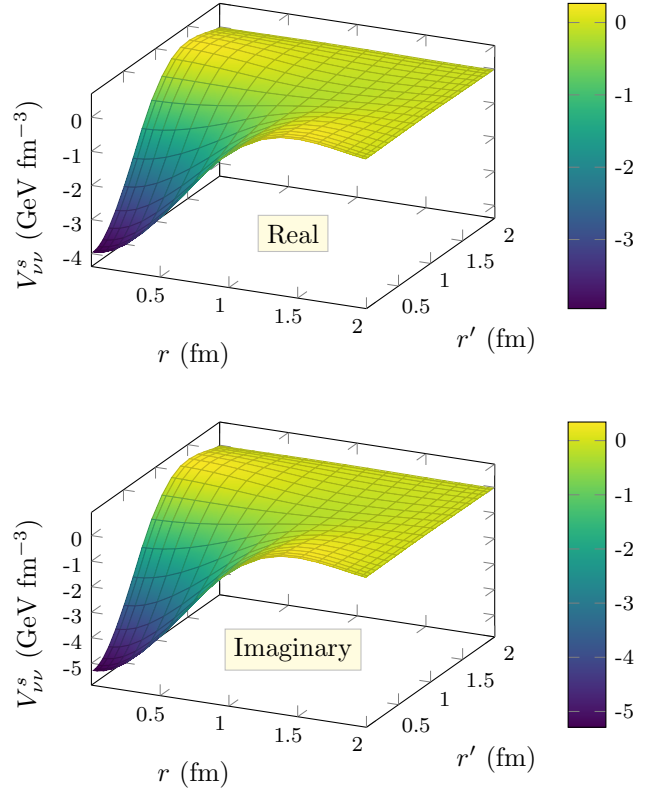
\begin{figure}
\centering
\begin{tikzpicture}
  \begin{groupplot}[
    group style={
      group size=1 by 2,
      horizontal sep=0.0cm,
      vertical sep=1.5cm
    },
    xlabel={$r$ (fm)},
    xmin=0,
    xmax=2,
    xtick={0.5,1,1.5,2},
    xticklabels={0.5,1,1.5,2},
    ylabel={$r'$ (fm)},
    ymin=0,
    ymax=2,
    ytick={0.5,1,1.5,2},
    yticklabels={0.5,1,1.5,2},
    zlabel={$V^s_{\nu \nu}$ (GeV fm${}^{-3}$)},
    tick label style={font=\footnotesize},
 width=0.8\columnwidth,
height=0.65\columnwidth,
  ]

    \nextgroupplot[
      colorbar,
      colorbar style={
        ytick={-4000,-3000,-2000,-1000,0},
        yticklabels={-4,-3,-2,-1,0},
        xshift=-0.0cm
      },
      ztick={-4000,-3000,-2000,-1000,0},
      zticklabels={-4,-3,-2,-1,0},
      title={Real},
      title style={
        fill=yellow!15!white,
        draw=black!25,
        font=\normalsize,
        yshift=-3.5cm,
        xshift=0cm
      }
    ]
      \addplot3[surf, shader=faceted interp, domain=0:10]
        file {kernel_nnb_real_he4_j1_pi-_nocoul_n20_noreg.dat};

    \nextgroupplot[
      colorbar,
      colorbar style={
        ytick={-5000,-4000,-3000,-2000,-1000,0},
        yticklabels={-5,-4,-3,-2,-1,0},
        xshift=-0.0cm
      },
      ztick={-5000,-4000,-3000,-2000,-1000,0},
      zticklabels={-5,-4,-3,-2,-1,0},
      title={Imaginary},
      title style={
        fill=yellow!15!white,
        draw=black!25,
        font=\normalsize,
        yshift=-3.5cm,
        xshift=0cm
      }
    ]
      \addplot3[surf, shader=faceted interp, domain=0:10]
        file {kernel_nnb_im_he4_j1_pi-_nocoul_n20_noreg.dat};

  \end{groupplot}
\end{tikzpicture}
\caption{Real (top) and imaginary (bottom) parts of the antinucleon-${}^4 \rm He$ strong-interaction potential kernel $V^s_{\nu \nu}(r',r)$ (in GeV fm${}^{-3}$) for the ${}^2S_{1/2}^-$ channel, extracted from the NCSM/RGM calculation with $N_{\text{max}}=20$. The kernel is shown in the interval $r,r' \leq 2$ fm, and is negligible elsewhere. The calculation uses the ground state of the target in the NCSM/RGM ansatz [\autoref{eq:rgm_ansatz_integral_form}].}
\label{fig:kern_alpha_-}
\end{figure}

We now focus on the system of interest in this work, $\bar{p}$-${}^4 \rm He$. Unless stated otherwise, all calculations include only the target ground state. For this system, there is a single $s$-wave channel, (${}^2S_{1/2}^-$). The corresponding strong-interaction kernel is shown in \autoref{fig:kern_alpha_-}. A key feature is its compact range: the kernel is already negligible by $r,r'\approx 2$ fm. For direct comparison with the lighter targets studied in Ref.~\cite{paperprc}, we show the kernel at the same truncation,  $N_{\text{max}}=20$. The $\bar{p}$-${}^4 \rm He$ potential well is substantially deeper than for the lighter systems, which makes the numerics more challenging. At the same time, the tightly bound $\alpha$ particle is better suited to the present NCSM/RGM cluster expansion, which retains a single Jacobi set of the five-body coordinate system (see the discussion in Ref.~\cite{paperprc}). \par
%
%
%
%
Before moving on to the discussion of annihilation densities, we establish the validity of our approach by comparing our results with the available experimental data. These include the reaction and elastic differential cross section and antiprotonic atom quasibound states. Other useful observables for this system are provided in the Appendix. \par
In \autoref{fig:sigma_reac_pbar_alpha}, we show our results for the reaction cross section and compare it with the available experimental data. The calculation includes contributions from the $J=1/2$ and $J=3/2$ partial waves with both parities, which is sufficient for this energy range. Our results appear to be consistent with the available experimental data, although the experimental constraints remain scarce. Nevertheless, at $p_{\text{lab}}=192.8$ MeV, corresponding to $E_{\mathrm{c.m.}}=15.83$ MeV in non-relativistic kinematics, enough experimental points are available to enable a relevant comparison with the elastic differential cross section. This is shown in \autoref{fig:dsigma_pbar_alpha}. To reach convergence, we include contributions from $J=1/2$ up to $J=7/2$, with both parities. Apart from the angular region where the calculated curve reaches its minimum, our results show reasonable agreement with experiment. \par
%
%
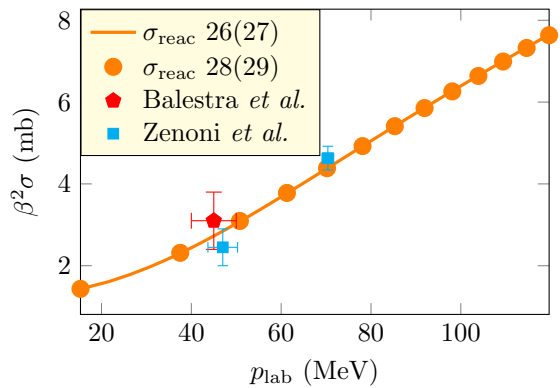
\begin{figure}
\centering
\begin{tikzpicture}
    \begin{groupplot}[
      group style={
        group size=1 by 1,
        horizontal sep=0.0cm,
        vertical sep=0.0cm
      },
      enlarge x limits=false,
             width=0.90\columnwidth,
height=0.65\columnwidth,
      clip mode=individual,
      legend style={
        fill=yellow!15!white,
        at={(0,1)},anchor=north west,
        nodes={scale=1, transform shape}
      }
    ]

      \nextgroupplot[
        legend cell align=left,
        xlabel={$p_{\text{lab}}$ (MeV)},
        ylabel={$\beta^2 \sigma$ (mb)}
      ]

        \addplot[very thick, mark size=3, color=orange]
          table[
            x expr=sqrt(\thisrowno{0}*2347.2963115675),
            y expr=2.662637848*\thisrowno{0}*\thisrowno{1}
          ] {sigma_reac_4h_p_j13_pi-+_n26_28_jr5_nosrg_hw20_mv1.dat};
        \addlegendentry{$\sigma_{\text{reac}}$ 26(27)}

        \addplot[only marks, thick, color=orange, mark size=3, mark repeat={5}]
          table[
            x expr=sqrt(\thisrowno{0}*2347.2963115675),
            y expr=2.662637848*\thisrowno{0}*\thisrowno{2}
          ] {sigma_reac_4h_p_j13_pi-+_n26_28_jr5_nosrg_hw20_mv1.dat};
        \addlegendentry{$\sigma_{\text{reac}}$ 28(29)}

        \addplot+[
          only marks,
          color=red,
          mark=pentagon*,
          mark size=3,
          mark options={fill=red},
          error bars/.cd,
            y dir=both, y explicit,
            x dir=both, x explicit
        ] coordinates {
          (45,3.1) +- (5,0.7)
        };
        \addlegendentry{Balestra \textit{et al.}}

        \addplot+[
          only marks,
          color=cyan,
          mark=square*,
          mark size=2,
          mark options={fill=cyan},
          error bars/.cd,
            y dir=both, y explicit,
            x dir=both, x explicit
        ] coordinates {
          (47,2.45)   +- (3.3,0.45)
          (70.4,4.63) +- (1.3,0.29)
        };
        \addlegendentry{Zenoni \textit{et al.}}

    \end{groupplot}
\end{tikzpicture}
\caption{$\bar{p}$-${}^4 \rm He$ reaction cross sections calculated with the regulator parameters $r_{\text{reg}}=7$ fm and $r_{\text{reg,c}}=5$ fm. The cross section is multiplied by the square of the antiproton beam's velocity in the laboratory frame, denoted with $\beta$. The convergence with respect to $N_{\text{max}}$ is demonstrated by comparing the results with $N_{\text{max}}=26(27)$ (line) and $N_{\text{max}}=28(29)$ (filled circles). The values inside the parentheses in the plot legend denote $N_{\text{max}}$ for the positive-parity states. The experimental values (squares and pentagon) are from Refs.~\cite{balestra1989antiproton,zenoni1999pd}.}
\label{fig:sigma_reac_pbar_alpha}
\end{figure}
%
%
\begin{figure}
\centering
\begin{tikzpicture}
    \begin{groupplot}[
    group style={
      group size=1 by 1,
      horizontal sep=0.0cm,
      vertical sep=0.0cm
    },
    ymode=log,
    enlarge x limits=false,
         width=0.90\columnwidth,
height=0.65\columnwidth,
    clip mode=individual
    ]

    \nextgroupplot[
      legend cell align=left,
      legend pos=north east,
      xlabel={$\,\theta_{\text{c.m.}}$},
      ylabel={$\dfrac{\mathrm{d}\sigma}{\mathrm{d}\Omega}$ (mb/sr)},
      legend style={at={(1,1)}, fill=yellow!15!white}
    ]

      \addplot[very thick, mark size=3, color=orange]
        table[x index=0, y index=1] {dsigma_dOmega_he4_nmax22_j1357_mv1_jrelm5.dat};
      \addlegendentry{$N_{\text{max}}=22(23)$}

      \addplot[only marks, mark size=3, color=cyan, mark repeat={5}]
        table[x index=0, y index=1] {dsigma_dOmega_he4_nmax24_j1357_mv1_jrelm5.dat};
      \addlegendentry{$N_{\text{max}}=24(25)$}

      \addplot+[
        only marks,
        color=red,
        mark=pentagon*,
        mark size=3,
        mark options={fill=red},
        error bars/.cd,
          y dir=both, y explicit
      ] coordinates {
        (14.2,124.94)  +- (0,15.15)
        (19.5,92.49)   +- (0,10.21)
        (24.5,74.69)   +- (0,8.20)
        (29.3,74.89)   +- (0,7.27)
        (34.5,65.25)   +- (0,6.22)
        (39.5,51.80)   +- (0,5.23)
        (44.6,51.23)   +- (0,4.95)
        (49.4,35.86)   +- (0,3.98)
        (54.6,34.25)   +- (0,3.74)
        (59.3,16.63)   +- (0,2.54)
        (64.2,13.30)   +- (0,2.22)
        (68.7,9.28)    +- (0,1.82)
        (74.7,4.14)    +- (0,1.19)
        (82.3,2.01)    +- (0,0.58)
        (94.1,1.17)    +- (0,0.44)
        (105.6,0.52)   +- (0,0.3)
        (111.5,0.36)   +- (0,0.25)
        (122.8,1.19)   +- (0,0.49)
        (132.3,1.58)   +- (0,0.6)
        (144.8,0.87)   +- (0,0.5)
        (153.0,1.10)   +- (0,0.63)
      };
      \addlegendentry{Experiment}

    \end{groupplot}
\end{tikzpicture}
\caption{$\bar{p}$-${}^4 \rm He$ elastic differential cross section evaluated at $E_{\mathrm{c.m.}}=15.83$ MeV with $N_{\text{max}}=22(23)$ (yellow line) and $N_{\text{max}}=24(25)$ (blue filled circles) in the center-of-mass frame. The regulator parameters are the same as in \autoref{fig:sigma_reac_pbar_alpha}. The experimental values (red pentagons) are from Ref.~\cite{BALESTRA199318}.}
\label{fig:dsigma_pbar_alpha}
\end{figure}
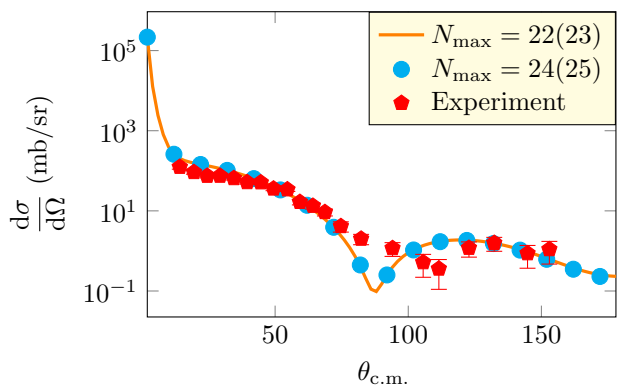
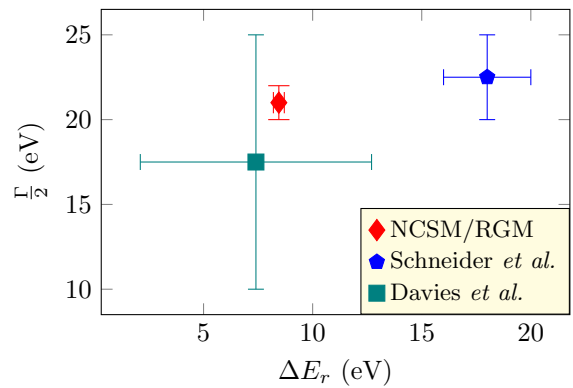
\begin{figure}[!h]
\centering
\begin{tikzpicture}
    \begin{axis}[
    xlabel={$\Delta E_r$ (eV)},
    ylabel={$\frac{\Gamma}{2}$ (eV)},
    width=0.90\columnwidth,
height=0.65\columnwidth,
    legend style={
      at={(1,0)},
      anchor=south east,
      fill=yellow!15!white,
      legend cell align=left,
      font=\small,
      nodes={scale=1, transform shape}
    }
    ]

    \addplot+[
      only marks,
      mark=diamond*,
      color=red,
      mark size=4pt,
      mark options={fill=red},
     error bars/.cd,
        y dir=both, y explicit,
        x dir=both, x explicit
    ] coordinates {
      (8.45,21) +- (0.25,1)
    };
    \addlegendentry{NCSM/RGM}

    \addplot+[
      only marks,
      mark=pentagon*,
      color=blue,
      mark size=3pt,
      mark options={fill=blue},
      error bars/.cd,
        y dir=both, y explicit,
        x dir=both, x explicit
    ] coordinates {
      (18,22.5) +- (2,2.5)
    };
    \addlegendentry{Schneider \textit{et al.}}

    \addplot+[
      only marks,
      mark=square*,
      color=teal,
      mark size=3pt,
      mark options={fill=teal},
      error bars/.cd,
        y dir=both, y explicit,
        x dir=both, x explicit
    ] coordinates {
      (7.4,17.5) +- (5.3,7.5)
    };
    \addlegendentry{Davies \textit{et al.}}

    \end{axis}
\end{tikzpicture}
\caption{Comparison of our averaged $\bar{p}$-${}^4 \rm He$ atomic level shift ($\Delta E_r$) and half-width ($\Gamma/2$) (diamond) with the experimental values of Davies \textit{et al.}~\cite{davies1984measurement} (square) and Schneider \textit{et al.}~\cite{schneider1991x} (pentagon). Our results correspond to the averaged 2P atomic level shifts and half-widths of \autoref{tab:level_he4_pwave} with $N_{\text{max}}=30$. The uncertainty in our result is due to dependence on the introduced regulator parameters.} 
\label{fig:compar_shift_he4}
\end{figure}

%
%
Due to strong interaction between the antiproton and the nucleus, the atomic levels of antiprotonic atoms are shifted and broadened~\cite{Batty:1997zp}. Here we calculate the averaged 2P ($n=2, \ell=1$) level shifts and half-widths using the $R$-matrix method, and compare with two experimental values in \autoref{fig:compar_shift_he4}. The predicted half-width is consistent with both experiments, while the real part is consistent with the result of Davies \textit{et al.}~\cite{davies1984measurement} but differs by roughly a factor of two from that of Schneider \textit{et al.}~\cite{schneider1991x}. Given the uncertainty associated with the optical \NNb potential, the many-body method, and the precision required to resolve level shifts in the order of a few eV, this level of agreement is encouraging but should be interpreted with caution.\par 
%
%
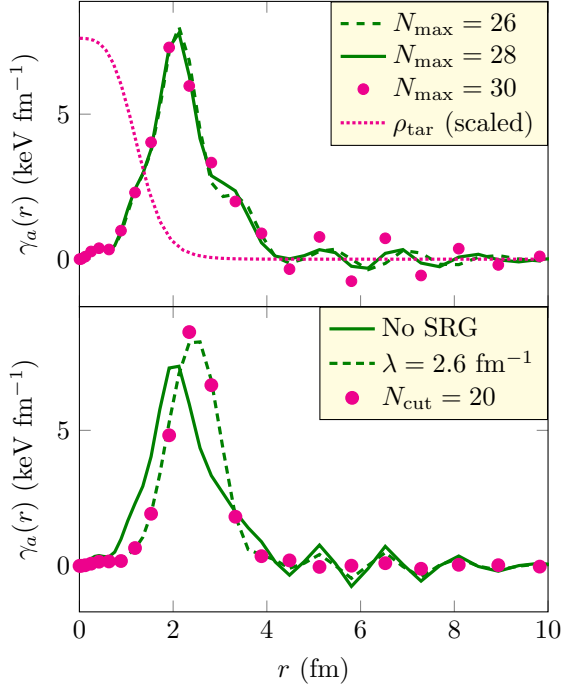
\begin{figure}
\centering
\begin{tikzpicture}
    \begin{groupplot}[
    xmin=0,
    xmax=10,
    ylabel={$\gamma_a(r)$ (keV fm${}^{-1}$)},
    group style={
      group size=1 by 2,
      horizontal sep=0cm,
      vertical sep=0cm
    },
    legend cell align=left,
        width=0.90\columnwidth,
height=0.65\columnwidth,
    legend style={
      at={(1,1)},
      fill=yellow!15!white,
      nodes={scale=1, transform shape}
    }
    ]

    \nextgroupplot[
      xticklabels=\empty
    ]

      \addplot[very thick, dashed, color=green!50!black]
        table[x index=0, y expr=\thisrowno{1}*1000]
        {annih_dens_he4_p_n_26_28_30_rreg7_rc5_ac200_ns300.dat};
      \addlegendentry{$N_{\text{max}}=26$}

      \addplot[very thick, color=green!50!black]
        table[x index=0, y expr=\thisrowno{2}*1000]
        {annih_dens_he4_p_n_26_28_30_rreg7_rc5_ac200_ns300.dat};
      \addlegendentry{$N_{\text{max}}=28$}

      \addplot[only marks, color=magenta, mark repeat=2]
        table[x index=0, y expr=\thisrowno{3}*1000]
        {annih_dens_he4_p_n_26_28_30_rreg7_rc5_ac200_ns300.dat};
      \addlegendentry{$N_{\text{max}}=30$}

      \addplot[very thick, magenta, densely dotted]
        table[x index=0, y expr=2*\thisrowno{1}]
        {targe_rho_4he_30_ns50_a10.dat};
      \addlegendentry{$\rho_{\text{tar}}$ (scaled)}

    \nextgroupplot[
      xlabel={$r$ (fm)}
    ]

      \addplot[very thick, color=green!50!black]
        table[x index=0, y expr=\thisrowno{3}*1000]
        {annih_dens_he4_p_n_26_28_30_rreg7_rc5_ac200_ns300.dat};
      \addlegendentry{No SRG}

      \addplot[very thick, color=green!50!black, densely dashed]
        table[x index=0, y expr=\thisrowno{1}*1000]
        {annih_density_n30_srg2.6_nocut.dat};
      \addlegendentry{$\lambda=2.6$ fm$^{-1}$}

      \addplot[very thick, only marks, mark repeat={2}, color=magenta]
        table[x index=0, y expr=\thisrowno{1}*1000]
        {annih_density_30_cut20_srg2.6.dat};
      \addlegendentry{$N_{\mathrm{cut}}=20$}

    \end{groupplot}
\end{tikzpicture}
\caption{$\bar{p}$-${}^4 \rm He$ annihilation density ($\gamma_a$) for the lowest atomic state in the ${}^2S^{-}_{1/2}$ channel as a function of antiproton's distance from the center-of-mass of the target ($r$). Top: Dependence on $N_{\text{max}}$, 26 (dashed), 28 (solid), and 30 (symbols). The dotted line denotes the target's density. Bottom: Comparison of the results obtained using the unevolved Hamiltonian (solid line) with those obtained using SRG with $\lambda=2.6$ fm${}^{-1}$ (dashed line) at $N_{\text{max}}=30$. The symbols denote the calculation with the same $N_{\text{max}}$ and $\lambda$, with the truncation of the SRG-evolved NCSM/RGM potential at $N_{\text{cut}}=20$ when calculating the density. We use $r_{\text{reg}}=7$ fm, $r_{\text{reg,c}}=5$ fm, $a_c=200$ fm and $n_s=300$. For the SRG curves, we use $r_{\text{reg}}=r_{\text{reg,c}}=7$ fm.}
\label{fig:p_he4_annih_density}
\end{figure}

%
%
We now turn to the central result of this work. Having established that the chosen Hamiltonian and the NCSM/RGM framework adapted to antiprotonic systems capture the main physics of the antiprotonic ${}^4 \mathrm{He}$ system at a qualitatively reliable (and, to some extent, quantitatively reliable) level, we focus on the key quantity relevant to experiments such as PUMA at CERN. In the top panel of \autoref{fig:p_he4_annih_density}, we show the annihilation density for the ground state of the antiprotonic atom in the ${}^2S^{-}_{1/2}$ channel. For comparison, we also show the target density computed from the translationally invariant NCSM wave function~\cite{PhysRevC.70.014317}. The annihilation density peaks in the tail of the target density ($r \approx 2$ fm), and is negligible for $r < 1$ fm. If the annihilation density is interpreted as characterizing the spatial distribution for the annihilation process, these results confirm the phenomenological picture that antiproton annihilation is predominantly peripheral. We also note that the qualitative features of the density are converged with respect to the model-space size, although small oscillations remain in its tail. These oscillations are numerical artifacts associated with the high-$n$ components of the HO-basis expansion of the kernels in \autoref{eq:direct_def}. While they are still visible here, our previous calculations for lighter targets indicate that such artifacts can be progressively reduced by increasing the model-space size~\cite{dehghani}. \par
To clarify the origin of the oscillation at the tail of the annihilation density, we apply a similarity renormalization group (SRG) transformation~\cite{bogner2007similarity,bogner2008three, Anderson:2010aq} to the $\bar{p}$-${}^4 \mathrm{He}$ Hamiltonian obtained in the NCSM/RGM basis. The purpose of the SRG evolution is to decouple the low- and high-$n$ components of the $\bar{p}$-${}^4 \mathrm{He}$ Hamiltonian, so that low-$n$ matrix elements can represent the essential features of the potential kernel. In this case, the sum in \autoref{eq:direct_def} can be reduced to the low-energy part of the basis.
This is shown in the bottom panel of \autoref{fig:p_he4_annih_density}, where the result obtained with the unevolved Hamiltonian (solid curve) is compared with that from the SRG-evolved Hamiltonian (dashed curve). We observe that, even after SRG evolution, the oscillations associated with high-$n$ HO wave functions persist. However, when the summation in \autoref{eq:direct_def} is restricted to $N_{\mathrm {cut}}=20$ (symbols), the SRG result preserves the same overall shape while the oscillatory tail disappears. \par
%
%
%
%
It is important to note that the SRG-based remedy does not necessarily preserve the detailed qualitative features of the annihilation density, since the annihilation density is not an observable. Nevertheless, once the low-energy content of the annihilation density is isolated, the annihilation strength is displaced towards larger intercluster distances. This reinforces the conclusion that annihilation is predominantly peripheral in antiprotonic systems, already from the $\alpha$ particle, which is much closer to a standard well-bound nucleus than the very light few-body systems.\par
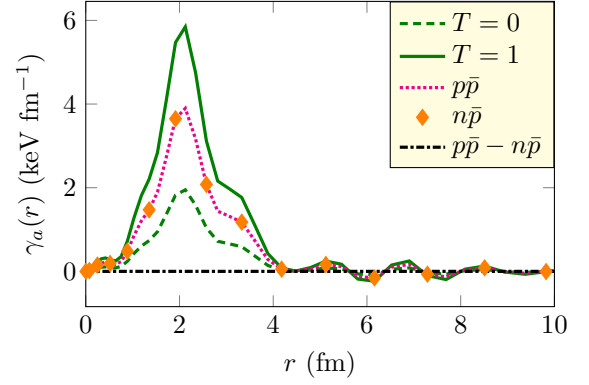
\begin{figure}
\centering
\begin{tikzpicture}
    \begin{groupplot}[
      xmin=0,
      xmax=10,
      xlabel={$r$ (fm)},
      ylabel={$\gamma_a(r)$ (keV fm${}^{-1}$)},
      group style={
        group size=1 by 1,
        horizontal sep=0cm,
        vertical sep=0cm
      },
      legend cell align=left,
         width=0.90\columnwidth,
height=0.65\columnwidth,
      legend style={at={(1,1)}, fill=yellow!15!white}
    ]
    
      \nextgroupplot

        \addplot[very thick, densely dashed, color=green!50!black]
          table[x index=0, y expr=\thisrowno{1}*1000]
          {annih_density_T0_T1_n28_rreg_7_5.dat};
        \addlegendentry{$T=0$}
    
        \addplot[very thick, color=green!50!black]
          table[x index=0, y expr=\thisrowno{2}*1000]
          {annih_density_T0_T1_n28_rreg_7_5.dat};
        \addlegendentry{$T=1$}
    
        \addplot[very thick, densely dotted, color=magenta]
          table[
            x index=0,
            y expr=\thisrowno{1}*1000 + 1/3*\thisrowno{2}*1000
          ] {annih_density_T0_T1_n28_rreg_7_5.dat};
        \addlegendentry{$p\bar{p}$}
    
        \addplot[only marks, color=orange, mark=diamond*, mark size=3, mark repeat={3}]
          table[
            x index=0,
            y expr=2/3*\thisrowno{2}*1000
          ] {annih_density_T0_T1_n28_rreg_7_5.dat};
        \addlegendentry{$n\bar{p}$}
    
        \addplot[very thick, densely dashdotted, color=black]
          table[
            x index=0,
            y expr=\thisrowno{1}*1000 + 1/3*\thisrowno{2}*1000 - 2/3*\thisrowno{2}*1000
          ] {annih_density_T0_T1_n28_rreg_7_5.dat};
        \addlegendentry{$p\bar{p}-n\bar{p}$}

    \end{groupplot}
\end{tikzpicture}
\caption{Contribution of $T=0$ (dashed) and $T=1$ (solid) \NNb two-body matrix elements to the total annihilation density, together with an estimation of the contribution of $p\bar{p}$ (dotted curve) and $n\bar{p}$ (diamonds). Due to the \NNb interaction used, the difference between these two contributions (dashdotted line) is zero. The results are calculated with $N_{\text{max}}=28$, while the rest of the parameters are the same as the top panel of \autoref{fig:p_he4_annih_density}.}
\label{fig:p_he4_annih_density_separated}
\end{figure}

To gain further insight into the annihilation process, we separate the contributions of the $T=0$ and $T=1$ \NNb two-body matrix elements to the annihilation density. Since $p\bar p$ couples to both $T=0$ and $T=1$ isospin states, whereas $n\bar p$ is a pure $T=1$ state, we can, as a first approximation, estimate $\gamma_{p\bar{p}} = \gamma_{0}+\frac{1}{3}\gamma_{1}$ and $\gamma_{n\bar{p}} = \frac{2}{3}\gamma_{1}$, where $\gamma_0$ and $\gamma_1$ denote the contribution due to $T=0$ and $T=1$ matrix elements, respectively.
The corresponding curves for $\gamma_{p\bar p}$ (dotted) and $\gamma_{n\bar p}$ (diamonds) give identical contributions to the total width. This is shown in \autoref{fig:p_he4_annih_density_separated}, where we also display the separate $T=0$ and $T=1$ contributions (dashed and solid curves).
One may be tempted to interpret this result as indicating an equal number of annihilations of protons and neutrons. The KW optical potential employs an imaginary part that is independent of energy and spin-isospin channel. A more quantitative extraction of proton and neutron annihilation components would therefore require a more involved annihilation potential. \par
%
%
%
%
In summary, we presented an \abinitio study of the $\bar{p}$-${}^4 \rm He$ system within the NCSM/RGM framework. After establishing that our results for different observables are consistent with the available experimental data, we showed that the microscopic calculation of the annihilation mechanism supports a peripheral annihilation process, a feature that is likely to persist in heavier targets. This confirms the phenomenological interpretation and supports one of the central hypotheses of the experiments using antiprotons as a probe of the nuclear surface.
\begin{acknowledgments}
AD and GH express their gratitude to Jaume Carbonell, S\l awomir Wycech, Pierre-Yves Duerinck, and Rimantas Lazauskas for insightful discussions and for providing benchmarks. This material is based in part upon work supported by the U.S. Department of Energy, Office of Science, Office of Nuclear Physics, under Work Proposal No. SCW0498. This work was prepared in part by LLNL under Contract No. DE-AC52-07NA27344. PN acknowledges support from the NSERC Grant No. SAPIN-2022-00019. TRIUMF receives federal funding via a contribution agreement with the National Research Council of Canada. This project was provided with computing HPC and storage resources by GENCI at IDRIS/TGCC, thanks to the grant 2015-0513012 on the supercomputer Jean Zay/Joliot Curie. GH gratefully acknowledges support from the CNRS/IN2P3 Computing Center (Lyon, France) for providing computing and data-processing resources needed for this work. GH and AD acknowledge the ANR-FRANCE (French National Research Agency) for its financial support of the grant No. ANR-21-CE31-0020.
\end{acknowledgments}

\bibliography{references}

\begin{thebibliography}{35}%
\makeatletter
\providecommand \@ifxundefined [1]{%
 \@ifx{#1\undefined}
}%
\providecommand \@ifnum [1]{%
 \ifnum #1\expandafter \@firstoftwo
 \else \expandafter \@secondoftwo
 \fi
}%
\providecommand \@ifx [1]{%
 \ifx #1\expandafter \@firstoftwo
 \else \expandafter \@secondoftwo
 \fi
}%
\providecommand \natexlab [1]{#1}%
\providecommand \enquote  [1]{``#1''}%
\providecommand \bibnamefont  [1]{#1}%
\providecommand \bibfnamefont [1]{#1}%
\providecommand \citenamefont [1]{#1}%
\providecommand \href@noop [0]{\@secondoftwo}%
\providecommand \href [0]{\begingroup \@sanitize@url \@href}%
\providecommand \@href[1]{\@@startlink{#1}\@@href}%
\providecommand \@@href[1]{\endgroup#1\@@endlink}%
\providecommand \@sanitize@url [0]{\catcode `\\12\catcode `\$12\catcode
  `\&12\catcode `\#12\catcode `\^12\catcode `\_12\catcode `\%12\relax}%
\providecommand \@@startlink[1]{}%
\providecommand \@@endlink[0]{}%
\providecommand \url  [0]{\begingroup\@sanitize@url \@url }%
\providecommand \@url [1]{\endgroup\@href {#1}{\urlprefix }}%
\providecommand \urlprefix  [0]{URL }%
\providecommand \Eprint [0]{\href }%
\providecommand \doibase [0]{https://doi.org/}%
\providecommand \selectlanguage [0]{\@gobble}%
\providecommand \bibinfo  [0]{\@secondoftwo}%
\providecommand \bibfield  [0]{\@secondoftwo}%
\providecommand \translation [1]{[#1]}%
\providecommand \BibitemOpen [0]{}%
\providecommand \bibitemStop [0]{}%
\providecommand \bibitemNoStop [0]{.\EOS\space}%
\providecommand \EOS [0]{\spacefactor3000\relax}%
\providecommand \BibitemShut  [1]{\csname bibitem#1\endcsname}%
\let\auto@bib@innerbib\@empty
\bibitem [{\citenamefont {Caravita}\ \emph {et~al.}(2025)\citenamefont
  {Caravita} \emph {et~al.}}]{caravita2025cern}%
  \BibitemOpen
  \bibfield  {author} {\bibinfo {author} {\bibfnamefont {R.}~\bibnamefont
  {Caravita}} \emph {et~al.},\ }\href@noop {} {\bibinfo {title} {{CERN AD/ELENA
  antimatter program}}} (\bibinfo {year} {2025}),\ \Eprint
  {https://arxiv.org/abs/2503.22471} {arXiv:2503.22471 [nucl-ex]} \BibitemShut
  {NoStop}%
\bibitem [{\citenamefont {Duerinck}\ and\ \citenamefont
  {Lazauskas}(2026)}]{Duerinck:2026otx}%
  \BibitemOpen
  \bibfield  {author} {\bibinfo {author} {\bibfnamefont {P.-Y.}\ \bibnamefont
  {Duerinck}}\ and\ \bibinfo {author} {\bibfnamefont {R.}~\bibnamefont
  {Lazauskas}},\ }\bibfield  {title} {\bibinfo {title} {{\textit{Ab initio}
  description of $\bar{p}+{}^{3} \mathrm{H}$ and $\bar{p}+{}^{3} \mathrm{He}$
  systems in optical models}},\ }\href {https://doi.org/10.1103/c7s6-x4v9}
  {\bibfield  {journal} {\bibinfo  {journal} {Phys. Rev. C}\ }\textbf {\bibinfo
  {volume} {113}},\ \bibinfo {pages} {054003} (\bibinfo {year} {2026})},\
  \Eprint {https://arxiv.org/abs/2601.06541} {arXiv:2601.06541 [nucl-th]}
  \BibitemShut {NoStop}%
\bibitem [{\citenamefont {Duerinck}\ \emph {et~al.}(2023)\citenamefont
  {Duerinck}, \citenamefont {Lazauskas},\ and\ \citenamefont
  {Dohet-Eraly}}]{duerinck2023antiproton}%
  \BibitemOpen
  \bibfield  {author} {\bibinfo {author} {\bibfnamefont {P.-Y.}\ \bibnamefont
  {Duerinck}}, \bibinfo {author} {\bibfnamefont {R.}~\bibnamefont
  {Lazauskas}},\ and\ \bibinfo {author} {\bibfnamefont {J.}~\bibnamefont
  {Dohet-Eraly}},\ }\bibfield  {title} {\bibinfo {title} {{Antiproton-deuteron
  hydrogenic states from a coupled-channel approach}},\ }\href
  {https://doi.org/10.1103/PhysRevC.108.054003} {\bibfield  {journal} {\bibinfo
   {journal} {Phys. Rev. C}\ }\textbf {\bibinfo {volume} {108}},\ \bibinfo
  {pages} {054003} (\bibinfo {year} {2023})}\BibitemShut {NoStop}%
\bibitem [{\citenamefont {Vorabbi}\ \emph {et~al.}(2020)\citenamefont
  {Vorabbi}, \citenamefont {Gennari}, \citenamefont {Finelli}, \citenamefont
  {Giusti},\ and\ \citenamefont {Navr{\'a}til}}]{Vorabbi:2019ciy}%
  \BibitemOpen
  \bibfield  {author} {\bibinfo {author} {\bibfnamefont {M.}~\bibnamefont
  {Vorabbi}}, \bibinfo {author} {\bibfnamefont {M.}~\bibnamefont {Gennari}},
  \bibinfo {author} {\bibfnamefont {P.}~\bibnamefont {Finelli}}, \bibinfo
  {author} {\bibfnamefont {C.}~\bibnamefont {Giusti}},\ and\ \bibinfo {author}
  {\bibfnamefont {P.}~\bibnamefont {Navr{\'a}til}},\ }\bibfield  {title}
  {\bibinfo {title} {Elastic antiproton-nucleus scattering from chiral
  forces},\ }\href {https://doi.org/10.1103/PhysRevLett.124.162501} {\bibfield
  {journal} {\bibinfo  {journal} {Phys. Rev. Lett.}\ }\textbf {\bibinfo
  {volume} {124}},\ \bibinfo {pages} {162501} (\bibinfo {year} {2020})},\
  \Eprint {https://arxiv.org/abs/1906.11984} {arXiv:1906.11984 [nucl-th]}
  \BibitemShut {NoStop}%
\bibitem [{\citenamefont {Aumann}\ \emph {et~al.}(2022)\citenamefont {Aumann}
  \emph {et~al.}}]{puma}%
  \BibitemOpen
  \bibfield  {author} {\bibinfo {author} {\bibfnamefont {T.}~\bibnamefont
  {Aumann}} \emph {et~al.} (\bibinfo {collaboration} {PUMA}),\ }\bibfield
  {title} {\bibinfo {title} {{PUMA, antiProton unstable matter annihilation}},\
  }\href {https://doi.org/10.1140/epja/s10050-022-00713-x} {\bibfield
  {journal} {\bibinfo  {journal} {Eur. Phys. J. A}\ }\textbf {\bibinfo {volume}
  {58}},\ \bibinfo {pages} {88} (\bibinfo {year} {2022})}\BibitemShut {NoStop}%
\bibitem [{\citenamefont {Trzcinska}\ \emph {et~al.}(2001)\citenamefont
  {Trzcinska} \emph {et~al.}}]{Trzcinska:2001sy}%
  \BibitemOpen
  \bibfield  {author} {\bibinfo {author} {\bibfnamefont {A.}~\bibnamefont
  {Trzcinska}} \emph {et~al.},\ }\bibfield  {title} {\bibinfo {title} {{Neutron
  density distributions deduced from antiprotonic atoms}},\ }\href
  {https://doi.org/10.1103/PhysRevLett.87.082501} {\bibfield  {journal}
  {\bibinfo  {journal} {Phys. Rev. Lett.}\ }\textbf {\bibinfo {volume} {87}},\
  \bibinfo {pages} {082501} (\bibinfo {year} {2001})}\BibitemShut {NoStop}%
\bibitem [{\citenamefont {Lubinski}\ \emph {et~al.}(1998)\citenamefont
  {Lubinski} \emph {et~al.}}]{Lubinski:1998xf}%
  \BibitemOpen
  \bibfield  {author} {\bibinfo {author} {\bibfnamefont {P.}~\bibnamefont
  {Lubinski}} \emph {et~al.},\ }\bibfield  {title} {\bibinfo {title}
  {{Composition of the nuclear periphery from antiproton absorption}},\ }\href
  {https://doi.org/10.1103/PhysRevC.57.2962} {\bibfield  {journal} {\bibinfo
  {journal} {Phys. Rev. C}\ }\textbf {\bibinfo {volume} {57}},\ \bibinfo
  {pages} {2962} (\bibinfo {year} {1998})},\ \Eprint
  {https://arxiv.org/abs/nucl-ex/9808005} {arXiv:nucl-ex/9808005} \BibitemShut
  {NoStop}%
\bibitem [{\citenamefont {Barrett}\ \emph {et~al.}(2013)\citenamefont
  {Barrett}, \citenamefont {Navr{\'a}til},\ and\ \citenamefont
  {Vary}}]{Barrett:2013nh}%
  \BibitemOpen
  \bibfield  {author} {\bibinfo {author} {\bibfnamefont {B.~R.}\ \bibnamefont
  {Barrett}}, \bibinfo {author} {\bibfnamefont {P.}~\bibnamefont
  {Navr{\'a}til}},\ and\ \bibinfo {author} {\bibfnamefont {J.~P.}\ \bibnamefont
  {Vary}},\ }\bibfield  {title} {\bibinfo {title} {{\textit{Ab initio} no core
  shell model}},\ }\href {https://doi.org/10.1016/j.ppnp.2012.10.003}
  {\bibfield  {journal} {\bibinfo  {journal} {Prog. Part. Nucl. Phys.}\
  }\textbf {\bibinfo {volume} {69}},\ \bibinfo {pages} {131} (\bibinfo {year}
  {2013})}\BibitemShut {NoStop}%
\bibitem [{\citenamefont {Navr{\'a}til}\ \emph {et~al.}(2000)\citenamefont
  {Navr{\'a}til}, \citenamefont {Kamuntavicius},\ and\ \citenamefont
  {Barrett}}]{Navr_til_2000}%
  \BibitemOpen
  \bibfield  {author} {\bibinfo {author} {\bibfnamefont {P.}~\bibnamefont
  {Navr{\'a}til}}, \bibinfo {author} {\bibfnamefont {G.~P.}\ \bibnamefont
  {Kamuntavicius}},\ and\ \bibinfo {author} {\bibfnamefont {B.~R.}\
  \bibnamefont {Barrett}},\ }\bibfield  {title} {\bibinfo {title} {{Few nucleon
  systems in translationally invariant harmonic oscillator basis}},\ }\href
  {https://doi.org/10.1103/PhysRevC.61.044001} {\bibfield  {journal} {\bibinfo
  {journal} {Phys. Rev. C}\ }\textbf {\bibinfo {volume} {61}},\ \bibinfo
  {pages} {044001} (\bibinfo {year} {2000})},\ \Eprint
  {https://arxiv.org/abs/nucl-th/9907054} {arXiv:nucl-th/9907054} \BibitemShut
  {NoStop}%
\bibitem [{\citenamefont {Navratil}(2007)}]{navratil2008ab}%
  \BibitemOpen
  \bibfield  {author} {\bibinfo {author} {\bibfnamefont {P.}~\bibnamefont
  {Navratil}},\ }\bibfield  {title} {\bibinfo {title} {{\textit{Ab initio}
  no-core shell model calculations for light nuclei}},\ }in\ \href@noop {}
  {\emph {\bibinfo {booktitle} {{169th Course of International School of
  Physics `Enrico Fermi'}: {Nuclear Structure far from Stability: New Physics
  and New Technology}}}}\ (\bibinfo  {publisher} {IOS, Amsterdam},\ \bibinfo
  {year} {2007})\ pp.\ \bibinfo {pages} {147--183},\ \Eprint
  {https://arxiv.org/abs/0711.2702} {arXiv:0711.2702 [nucl-th]} \BibitemShut
  {NoStop}%
\bibitem [{\citenamefont {Quaglioni}\ and\ \citenamefont
  {Navr{\'a}til}(2008)}]{Quaglioni:2008sm}%
  \BibitemOpen
  \bibfield  {author} {\bibinfo {author} {\bibfnamefont {S.}~\bibnamefont
  {Quaglioni}}\ and\ \bibinfo {author} {\bibfnamefont {P.}~\bibnamefont
  {Navr{\'a}til}},\ }\bibfield  {title} {\bibinfo {title} {{\textit{Ab initio}
  many-body calculations of $n$-${}^3${H}, $n$-${}^4${H}e, $p$-${}^{3,4}${H}e
  and $n$-${}^{10}${B}e scattering}},\ }\href
  {https://doi.org/10.1103/PhysRevLett.101.092501} {\bibfield  {journal}
  {\bibinfo  {journal} {Phys. Rev. Lett.}\ }\textbf {\bibinfo {volume} {101}},\
  \bibinfo {pages} {092501} (\bibinfo {year} {2008})},\ \Eprint
  {https://arxiv.org/abs/0804.1560} {arXiv:0804.1560 [nucl-th]} \BibitemShut
  {NoStop}%
\bibitem [{\citenamefont {Quaglioni}\ and\ \citenamefont
  {Navr{\'a}til}(2009)}]{quaglioni}%
  \BibitemOpen
  \bibfield  {author} {\bibinfo {author} {\bibfnamefont {S.}~\bibnamefont
  {Quaglioni}}\ and\ \bibinfo {author} {\bibfnamefont {P.}~\bibnamefont
  {Navr{\'a}til}},\ }\bibfield  {title} {\bibinfo {title} {{\textit{Ab initio}
  many-body calculations of nucleon-nucleus scattering}},\ }\href
  {https://doi.org/10.1103/PhysRevC.79.044606} {\bibfield  {journal} {\bibinfo
  {journal} {Phys. Rev. C}\ }\textbf {\bibinfo {volume} {79}},\ \bibinfo
  {pages} {044606} (\bibinfo {year} {2009})},\ \Eprint
  {https://arxiv.org/abs/0901.0950} {arXiv:0901.0950 [nucl-th]} \BibitemShut
  {NoStop}%
\bibitem [{\citenamefont {Navr{\'a}til}\ \emph {et~al.}(2016)\citenamefont
  {Navr{\'a}til}, \citenamefont {Quaglioni}, \citenamefont {Hupin},
  \citenamefont {Romero-Redondo},\ and\ \citenamefont {Calci}}]{unified}%
  \BibitemOpen
  \bibfield  {author} {\bibinfo {author} {\bibfnamefont {P.}~\bibnamefont
  {Navr{\'a}til}}, \bibinfo {author} {\bibfnamefont {S.}~\bibnamefont
  {Quaglioni}}, \bibinfo {author} {\bibfnamefont {G.}~\bibnamefont {Hupin}},
  \bibinfo {author} {\bibfnamefont {C.}~\bibnamefont {Romero-Redondo}},\ and\
  \bibinfo {author} {\bibfnamefont {A.}~\bibnamefont {Calci}},\ }\bibfield
  {title} {\bibinfo {title} {{Unified \textit{ab initio} approaches to nuclear
  structure and reactions}},\ }\href
  {https://doi.org/10.1088/0031-8949/91/5/053002} {\bibfield  {journal}
  {\bibinfo  {journal} {Phys. Scripta}\ }\textbf {\bibinfo {volume} {91}},\
  \bibinfo {pages} {053002} (\bibinfo {year} {2016})},\ \Eprint
  {https://arxiv.org/abs/1601.03765} {arXiv:1601.03765 [nucl-th]} \BibitemShut
  {NoStop}%
\bibitem [{\citenamefont {Dehghani}\ \emph {et~al.}(2026)\citenamefont
  {Dehghani}, \citenamefont {Hupin}, \citenamefont {Quaglioni},\ and\
  \citenamefont {Navr{\'a}til}}]{paperprc}%
  \BibitemOpen
  \bibfield  {author} {\bibinfo {author} {\bibfnamefont {A.}~\bibnamefont
  {Dehghani}}, \bibinfo {author} {\bibfnamefont {G.}~\bibnamefont {Hupin}},
  \bibinfo {author} {\bibfnamefont {S.}~\bibnamefont {Quaglioni}},\ and\
  \bibinfo {author} {\bibfnamefont {P.}~\bibnamefont {Navr{\'a}til}},\
  }\bibfield  {title} {\bibinfo {title} {{Light antiproton-nucleus systems at
  low energies with the \textit{ab initio} NCSM/RGM method}},\ }\href
  {https://doi.org/10.1103/ldr8-xfp1} {\bibfield  {journal} {\bibinfo
  {journal} {Phys. Rev. C}\ }\textbf {\bibinfo {volume} {114}},\ \bibinfo
  {pages} {014613} (\bibinfo {year} {2026})},\ \Eprint
  {https://arxiv.org/abs/2602.18162} {arXiv:2602.18162 [nucl-th]} \BibitemShut
  {NoStop}%
\bibitem [{\citenamefont {Descouvemont}\ and\ \citenamefont
  {Baye}(2010)}]{descouvemont2010r}%
  \BibitemOpen
  \bibfield  {author} {\bibinfo {author} {\bibfnamefont {P.}~\bibnamefont
  {Descouvemont}}\ and\ \bibinfo {author} {\bibfnamefont {D.}~\bibnamefont
  {Baye}},\ }\bibfield  {title} {\bibinfo {title} {{The $R$-matrix theory}},\
  }\href {https://doi.org/10.1088/0034-4885/73/3/036301} {\bibfield  {journal}
  {\bibinfo  {journal} {Rept. Prog. Phys.}\ }\textbf {\bibinfo {volume} {73}},\
  \bibinfo {pages} {036301} (\bibinfo {year} {2010})},\ \Eprint
  {https://arxiv.org/abs/1001.0678} {arXiv:1001.0678 [nucl-th]} \BibitemShut
  {NoStop}%
\bibitem [{\citenamefont {Hesse}\ \emph {et~al.}(1998)\citenamefont {Hesse},
  \citenamefont {Sparenberg}, \citenamefont {Van~Raemdonck},\ and\
  \citenamefont {Baye}}]{hesse1998coupled}%
  \BibitemOpen
  \bibfield  {author} {\bibinfo {author} {\bibfnamefont {M.}~\bibnamefont
  {Hesse}}, \bibinfo {author} {\bibfnamefont {J.~M.}\ \bibnamefont
  {Sparenberg}}, \bibinfo {author} {\bibfnamefont {F.}~\bibnamefont
  {Van~Raemdonck}},\ and\ \bibinfo {author} {\bibfnamefont {D.}~\bibnamefont
  {Baye}},\ }\bibfield  {title} {\bibinfo {title} {{Coupled-channel $R$-matrix
  method on a Lagrange mesh}},\ }\href
  {https://doi.org/10.1016/S0375-9474(98)00435-7} {\bibfield  {journal}
  {\bibinfo  {journal} {Nucl. Phys. A}\ }\textbf {\bibinfo {volume} {640}},\
  \bibinfo {pages} {37} (\bibinfo {year} {1998})}\BibitemShut {NoStop}%
\bibitem [{\citenamefont {Entem}\ and\ \citenamefont
  {Machleidt}(2003)}]{entem2003accurate}%
  \BibitemOpen
  \bibfield  {author} {\bibinfo {author} {\bibfnamefont {D.~R.}\ \bibnamefont
  {Entem}}\ and\ \bibinfo {author} {\bibfnamefont {R.}~\bibnamefont
  {Machleidt}},\ }\bibfield  {title} {\bibinfo {title} {{Accurate charge
  dependent nucleon nucleon potential at fourth order of chiral perturbation
  theory}},\ }\href {https://doi.org/10.1103/PhysRevC.68.041001} {\bibfield
  {journal} {\bibinfo  {journal} {Phys. Rev. C}\ }\textbf {\bibinfo {volume}
  {68}},\ \bibinfo {pages} {041001} (\bibinfo {year} {2003})},\ \Eprint
  {https://arxiv.org/abs/nucl-th/0304018} {arXiv:nucl-th/0304018} \BibitemShut
  {NoStop}%
\bibitem [{\citenamefont {Kohno}\ and\ \citenamefont
  {Weise}(1986)}]{kohno1986proton}%
  \BibitemOpen
  \bibfield  {author} {\bibinfo {author} {\bibfnamefont {M.}~\bibnamefont
  {Kohno}}\ and\ \bibinfo {author} {\bibfnamefont {W.}~\bibnamefont {Weise}},\
  }\bibfield  {title} {\bibinfo {title} {Proton-antiproton scattering and
  annihilation into two mesons},\ }\href
  {https://doi.org/10.1016/0375-9474(86)90098-9} {\bibfield  {journal}
  {\bibinfo  {journal} {Nucl. Phys. A}\ }\textbf {\bibinfo {volume} {454}},\
  \bibinfo {pages} {429} (\bibinfo {year} {1986})}\BibitemShut {NoStop}%
\bibitem [{\citenamefont {Ueda}(1979)}]{ueda1979antinucleon}%
  \BibitemOpen
  \bibfield  {author} {\bibinfo {author} {\bibfnamefont {T.}~\bibnamefont
  {Ueda}},\ }\bibfield  {title} {\bibinfo {title} {Antinucleon-nucleon
  potentials for bound and scattering states},\ }\href
  {https://doi.org/10.1143/PTP.62.1670} {\bibfield  {journal} {\bibinfo
  {journal} {Prog. Theor. Phys.}\ }\textbf {\bibinfo {volume} {62}},\ \bibinfo
  {pages} {1670} (\bibinfo {year} {1979})}\BibitemShut {NoStop}%
\bibitem [{\citenamefont {Richard}(2020)}]{richard2020antiproton}%
  \BibitemOpen
  \bibfield  {author} {\bibinfo {author} {\bibfnamefont {J.-M.}\ \bibnamefont
  {Richard}},\ }\bibfield  {title} {\bibinfo {title} {{Antiproton physics}},\
  }\href {https://doi.org/10.3389/fphy.2020.00006} {\bibfield  {journal}
  {\bibinfo  {journal} {Front. in Phys.}\ }\textbf {\bibinfo {volume} {8}},\
  \bibinfo {pages} {6} (\bibinfo {year} {2020})},\ \Eprint
  {https://arxiv.org/abs/1912.07385} {arXiv:1912.07385 [nucl-th]} \BibitemShut
  {NoStop}%
\bibitem [{\citenamefont {Carbonell}\ \emph {et~al.}(2023)\citenamefont
  {Carbonell}, \citenamefont {Hupin},\ and\ \citenamefont
  {Wycech}}]{carbonell2023comparison}%
  \BibitemOpen
  \bibfield  {author} {\bibinfo {author} {\bibfnamefont {J.}~\bibnamefont
  {Carbonell}}, \bibinfo {author} {\bibfnamefont {G.}~\bibnamefont {Hupin}},\
  and\ \bibinfo {author} {\bibfnamefont {S.}~\bibnamefont {Wycech}},\
  }\bibfield  {title} {\bibinfo {title} {{Comparison of $\bar{N}N$ optical
  models}},\ }\href {https://doi.org/10.1140/epja/s10050-023-01161-x}
  {\bibfield  {journal} {\bibinfo  {journal} {Eur. Phys. J. A}\ }\textbf
  {\bibinfo {volume} {59}},\ \bibinfo {pages} {259} (\bibinfo {year} {2023})},\
  \Eprint {https://arxiv.org/abs/2309.14831} {arXiv:2309.14831 [nucl-th]}
  \BibitemShut {NoStop}%
\bibitem [{\citenamefont {Balestra}\ \emph {et~al.}(1989)\citenamefont
  {Balestra} \emph {et~al.}}]{balestra1989antiproton}%
  \BibitemOpen
  \bibfield  {author} {\bibinfo {author} {\bibfnamefont {F.}~\bibnamefont
  {Balestra}} \emph {et~al.},\ }\bibfield  {title} {\bibinfo {title}
  {{Antiproton-helium annihilation around 44 MeV/$c$}},\ }\href
  {https://doi.org/10.1016/0370-2693(89)91649-3} {\bibfield  {journal}
  {\bibinfo  {journal} {Phys. Lett. B}\ }\textbf {\bibinfo {volume} {230}},\
  \bibinfo {pages} {36} (\bibinfo {year} {1989})}\BibitemShut {NoStop}%
\bibitem [{\citenamefont {Zenoni}\ \emph {et~al.}(1999)\citenamefont {Zenoni}
  \emph {et~al.}}]{zenoni1999pd}%
  \BibitemOpen
  \bibfield  {author} {\bibinfo {author} {\bibfnamefont {A.}~\bibnamefont
  {Zenoni}} \emph {et~al.},\ }\bibfield  {title} {\bibinfo {title} {{$\bar{p}D$
  and $\bar{p} {}^4 \mathrm{He}$ annihilation cross-sections at very
  low-energy}},\ }\href {https://doi.org/10.1016/S0370-2693(99)00890-4}
  {\bibfield  {journal} {\bibinfo  {journal} {Phys. Lett. B}\ }\textbf
  {\bibinfo {volume} {461}},\ \bibinfo {pages} {413} (\bibinfo {year}
  {1999})}\BibitemShut {NoStop}%
\bibitem [{\citenamefont {Balestra}\ \emph {et~al.}(1993)\citenamefont
  {Balestra} \emph {et~al.}}]{BALESTRA199318}%
  \BibitemOpen
  \bibfield  {author} {\bibinfo {author} {\bibfnamefont {F.}~\bibnamefont
  {Balestra}} \emph {et~al.},\ }\bibfield  {title} {\bibinfo {title}
  {{Antiproton-${}^4 \mathrm{He}$ interactions at 200 MeV/c}},\ }\href
  {https://doi.org/https://doi.org/10.1016/0370-2693(93)91099-9} {\bibfield
  {journal} {\bibinfo  {journal} {Phys. Lett. B}\ }\textbf {\bibinfo {volume}
  {305}},\ \bibinfo {pages} {18} (\bibinfo {year} {1993})}\BibitemShut
  {NoStop}%
\bibitem [{\citenamefont {Batty}\ \emph {et~al.}(1997)\citenamefont {Batty},
  \citenamefont {Friedman},\ and\ \citenamefont {Gal}}]{Batty:1997zp}%
  \BibitemOpen
  \bibfield  {author} {\bibinfo {author} {\bibfnamefont {C.~J.}\ \bibnamefont
  {Batty}}, \bibinfo {author} {\bibfnamefont {E.}~\bibnamefont {Friedman}},\
  and\ \bibinfo {author} {\bibfnamefont {A.}~\bibnamefont {Gal}},\ }\bibfield
  {title} {\bibinfo {title} {{Strong interaction physics from hadronic
  atoms}},\ }\href {https://doi.org/10.1016/S0370-1573(97)00011-2} {\bibfield
  {journal} {\bibinfo  {journal} {Phys. Rept.}\ }\textbf {\bibinfo {volume}
  {287}},\ \bibinfo {pages} {385} (\bibinfo {year} {1997})}\BibitemShut
  {NoStop}%
\bibitem [{\citenamefont {Davies}\ \emph {et~al.}(1984)\citenamefont {Davies}
  \emph {et~al.}}]{davies1984measurement}%
  \BibitemOpen
  \bibfield  {author} {\bibinfo {author} {\bibfnamefont {J.~D.}\ \bibnamefont
  {Davies}} \emph {et~al.},\ }\bibfield  {title} {\bibinfo {title}
  {{Measurement of strong interaction effects in antiprotonic helium atoms}},\
  }\href {https://doi.org/10.1016/0370-2693(84)90052-2} {\bibfield  {journal}
  {\bibinfo  {journal} {Phys. Lett. B}\ }\textbf {\bibinfo {volume} {145}},\
  \bibinfo {pages} {319} (\bibinfo {year} {1984})}\BibitemShut {NoStop}%
\bibitem [{\citenamefont {Schneider}\ \emph {et~al.}(1991)\citenamefont
  {Schneider} \emph {et~al.}}]{schneider1991x}%
  \BibitemOpen
  \bibfield  {author} {\bibinfo {author} {\bibfnamefont {M.}~\bibnamefont
  {Schneider}} \emph {et~al.},\ }\bibfield  {title} {\bibinfo {title} {{X-rays
  from antiprotonic ${}^3\mathrm{He}$ and ${}^4\mathrm{He}$}},\ }\href
  {https://doi.org/10.1007/BF01284797} {\bibfield  {journal} {\bibinfo
  {journal} {Z. Phys. A}\ }\textbf {\bibinfo {volume} {338}},\ \bibinfo {pages}
  {217} (\bibinfo {year} {1991})}\BibitemShut {NoStop}%
\bibitem [{\citenamefont {Navr\'atil}(2004)}]{PhysRevC.70.014317}%
  \BibitemOpen
  \bibfield  {author} {\bibinfo {author} {\bibfnamefont {P.}~\bibnamefont
  {Navr\'atil}},\ }\bibfield  {title} {\bibinfo {title} {{Translationally
  invariant density}},\ }\href {https://doi.org/10.1103/PhysRevC.70.014317}
  {\bibfield  {journal} {\bibinfo  {journal} {Phys. Rev. C}\ }\textbf {\bibinfo
  {volume} {70}},\ \bibinfo {pages} {014317} (\bibinfo {year}
  {2004})}\BibitemShut {NoStop}%
\bibitem [{\citenamefont {Dehghani}(2025)}]{dehghani}%
  \BibitemOpen
  \bibfield  {author} {\bibinfo {author} {\bibfnamefont {A.}~\bibnamefont
  {Dehghani}},\ }\emph {\bibinfo {title} {Reactions with antiprotons in the
  theory of cold nuclear collisions}},\ \href@noop {} {Ph.D. thesis},\ \bibinfo
   {school} {Paris-Saclay University} (\bibinfo {year} {2025}),\ \bibinfo
  {note}
  {\href{https://theses.hal.science/tel-05398592}{https://theses.hal.science/tel-05398592}}\BibitemShut
  {NoStop}%
\bibitem [{\citenamefont {Bogner}\ \emph {et~al.}(2007)\citenamefont {Bogner},
  \citenamefont {Furnstahl},\ and\ \citenamefont
  {Perry}}]{bogner2007similarity}%
  \BibitemOpen
  \bibfield  {author} {\bibinfo {author} {\bibfnamefont {S.~K.}\ \bibnamefont
  {Bogner}}, \bibinfo {author} {\bibfnamefont {R.~J.}\ \bibnamefont
  {Furnstahl}},\ and\ \bibinfo {author} {\bibfnamefont {R.~J.}\ \bibnamefont
  {Perry}},\ }\bibfield  {title} {\bibinfo {title} {Similarity renormalization
  group for nucleon-nucleon interactions},\ }\href
  {https://doi.org/10.1103/PhysRevC.75.061001} {\bibfield  {journal} {\bibinfo
  {journal} {Phys. Rev. C}\ }\textbf {\bibinfo {volume} {75}},\ \bibinfo
  {pages} {061001} (\bibinfo {year} {2007})}\BibitemShut {NoStop}%
\bibitem [{\citenamefont {Bogner}\ \emph {et~al.}(2008)\citenamefont {Bogner},
  \citenamefont {Furnstahl},\ and\ \citenamefont {Perry}}]{bogner2008three}%
  \BibitemOpen
  \bibfield  {author} {\bibinfo {author} {\bibfnamefont {S.}~\bibnamefont
  {Bogner}}, \bibinfo {author} {\bibfnamefont {R.}~\bibnamefont {Furnstahl}},\
  and\ \bibinfo {author} {\bibfnamefont {R.}~\bibnamefont {Perry}},\ }\bibfield
   {title} {\bibinfo {title} {Three-body forces produced by a similarity
  renormalization group transformation in a simple model},\ }\href
  {https://doi.org/https://doi.org/10.1016/j.aop.2007.09.001} {\bibfield
  {journal} {\bibinfo  {journal} {Ann. Phys.}\ }\textbf {\bibinfo {volume}
  {323}},\ \bibinfo {pages} {1478} (\bibinfo {year} {2008})}\BibitemShut
  {NoStop}%
\bibitem [{\citenamefont {Anderson}\ \emph {et~al.}(2010)\citenamefont
  {Anderson}, \citenamefont {Bogner}, \citenamefont {Furnstahl},\ and\
  \citenamefont {Perry}}]{Anderson:2010aq}%
  \BibitemOpen
  \bibfield  {author} {\bibinfo {author} {\bibfnamefont {E.~R.}\ \bibnamefont
  {Anderson}}, \bibinfo {author} {\bibfnamefont {S.~K.}\ \bibnamefont
  {Bogner}}, \bibinfo {author} {\bibfnamefont {R.~J.}\ \bibnamefont
  {Furnstahl}},\ and\ \bibinfo {author} {\bibfnamefont {R.~J.}\ \bibnamefont
  {Perry}},\ }\bibfield  {title} {\bibinfo {title} {Operator evolution via the
  similarity renormalization group {I}: {T}he deuteron},\ }\href
  {https://doi.org/10.1103/PhysRevC.82.054001} {\bibfield  {journal} {\bibinfo
  {journal} {Phys. Rev. C}\ }\textbf {\bibinfo {volume} {82}},\ \bibinfo
  {pages} {054001} (\bibinfo {year} {2010})},\ \Eprint
  {https://arxiv.org/abs/1008.1569} {arXiv:1008.1569 [nucl-th]} \BibitemShut
  {NoStop}%
\bibitem [{\citenamefont {Deser}\ \emph {et~al.}(1954)\citenamefont {Deser},
  \citenamefont {Goldberger}, \citenamefont {Baumann},\ and\ \citenamefont
  {Thirring}}]{Deser}%
  \BibitemOpen
  \bibfield  {author} {\bibinfo {author} {\bibfnamefont {S.}~\bibnamefont
  {Deser}}, \bibinfo {author} {\bibfnamefont {M.~L.}\ \bibnamefont
  {Goldberger}}, \bibinfo {author} {\bibfnamefont {K.}~\bibnamefont
  {Baumann}},\ and\ \bibinfo {author} {\bibfnamefont {W.}~\bibnamefont
  {Thirring}},\ }\bibfield  {title} {\bibinfo {title} {Energy level
  displacements in pi-mesonic atoms},\ }\href
  {https://doi.org/10.1103/PhysRev.96.774} {\bibfield  {journal} {\bibinfo
  {journal} {Phys. Rev.}\ }\textbf {\bibinfo {volume} {96}},\ \bibinfo {pages}
  {774} (\bibinfo {year} {1954})}\BibitemShut {NoStop}%
\bibitem [{\citenamefont {Trueman}(1961)}]{TRUEMAN196157}%
  \BibitemOpen
  \bibfield  {author} {\bibinfo {author} {\bibfnamefont {T.~L.}\ \bibnamefont
  {Trueman}},\ }\bibfield  {title} {\bibinfo {title} {Energy level shifts in
  atomic states of strongly-interacting particles},\ }\href
  {https://doi.org/10.1016/0029-5582(61)90115-8} {\bibfield  {journal}
  {\bibinfo  {journal} {Nucl. Phys.}\ }\textbf {\bibinfo {volume} {26}},\
  \bibinfo {pages} {57} (\bibinfo {year} {1961})}\BibitemShut {NoStop}%
\bibitem [{\citenamefont {Carbonell}\ \emph {et~al.}(1992)\citenamefont
  {Carbonell}, \citenamefont {Richard},\ and\ \citenamefont
  {Wycech}}]{Carbonell:1992wd}%
  \BibitemOpen
  \bibfield  {author} {\bibinfo {author} {\bibfnamefont {J.}~\bibnamefont
  {Carbonell}}, \bibinfo {author} {\bibfnamefont {J.-M.}\ \bibnamefont
  {Richard}},\ and\ \bibinfo {author} {\bibfnamefont {S.}~\bibnamefont
  {Wycech}},\ }\bibfield  {title} {\bibinfo {title} {{On the relation between
  protonium level shifts and nucleon-antinucleon scattering amplitudes}},\
  }\href {https://doi.org/10.1007/BF01291531} {\bibfield  {journal} {\bibinfo
  {journal} {Z. Phys. A}\ }\textbf {\bibinfo {volume} {343}},\ \bibinfo {pages}
  {325} (\bibinfo {year} {1992})}\BibitemShut {NoStop}%
\end{thebibliography}%

\appendix*

\section*{Appendix: Convergence tests for other observables} \label{appendix}
In this section, we provide our results for several useful observables for this system, including phase shifts, scattering length, and antiprotonic-atom quasibound states. These calculations are done within the $N_{\text{max}}=26-30$ range. The ground-state energies of the target for the corresponding $N^{\text{cluster}}_{\text{max}}=$26, 28, and 30 are $-$25.18, $-$25.27, and $-$25.31 MeV, respectively. The converged value for ${}^4 \rm He$ is $-$25.39 MeV~\cite{navratil2008ab}. \par
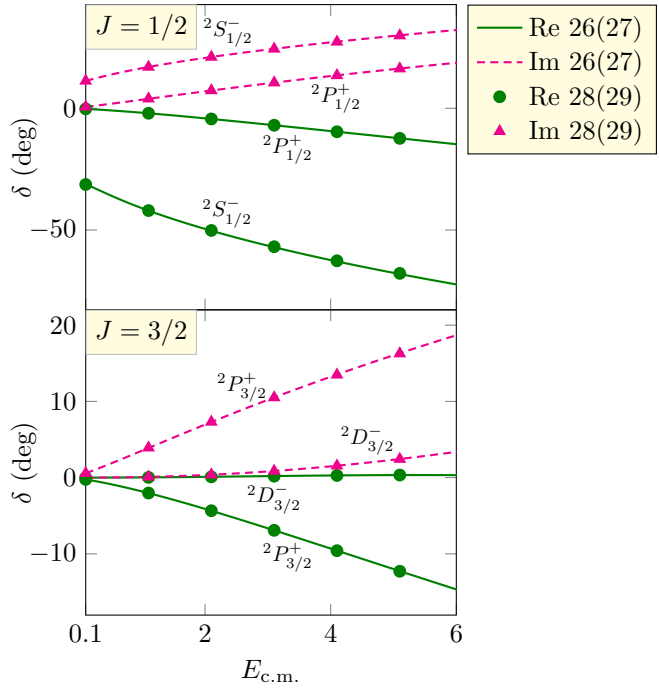
\begin{figure}[b]
\centering
\begin{tikzpicture}
    \begin{groupplot}[
    group style={
      group size=1 by 2,
      horizontal sep=1.5cm,
      vertical sep=0cm
    },
    width=0.75\columnwidth,
height=0.65\columnwidth,
    ylabel={$\delta$ (deg)},
    xtick={0.1,2,4,6},
    xticklabels={0.1,2,4,6},
    xmin=0.1,
    xmax=6,
    legend pos=outer north east,
    legend style={
      fill=yellow!15!white,
      legend cell align=left,
      nodes={scale=1, transform shape}
    },
    title style={
      fill=yellow!15!white,
      draw=black!25,
      at={(0.15,0.8)}
    }
    ]

    \nextgroupplot[
      title={$J=1/2$},
      ylabel style={yshift=-10pt},
      xticklabels=\empty
    ]

      \addplot[thick, green!50!black]
        table[x index=0, y index=1, col sep=space]
        {phase_shift_4h_p_j13_pi-+_n26_jr5_nosrg_hw20_mv1.dat}
        node[midway, above, black, scale=0.8] {${}^2S^{-}_{1/2}$};
      \addlegendentry{Re $26(27)$}

      \addplot[thick, magenta, densely dashed]
        table[x index=0, y index=1, col sep=space]
        {phase_shift_Im_4h_p_j13_pi-+_n26_jr5_nosrg_hw20_mv1.dat}
        node[midway, above, black, scale=0.8] {${}^2S^{-}_{1/2}$};
      \addlegendentry{Im $26(27)$}

      \addplot[thick, green!50!black, only marks, mark repeat={10}]
        table[x index=0, y index=1, col sep=space]
        {phase_shift_4h_p_j13_pi-+_n28_jr5_nosrg_hw20_mv1.dat};
      \addlegendentry{Re $28(29)$}

      \addplot[thick, magenta, only marks, mark repeat={10}, mark=triangle*]
        table[x index=0, y index=1, col sep=space]
        {phase_shift_Im_4h_p_j13_pi-+_n28_jr5_nosrg_hw20_mv1.dat};
      \addlegendentry{Im $28(29)$}

      \addplot[thick, green!50!black]
        table[x index=0, y index=2, col sep=space]
        {phase_shift_4h_p_j13_pi-+_n26_jr5_nosrg_hw20_mv1.dat}
        node[pos=0.5, below, black, scale=0.8] {${}^2P^{+}_{1/2}$};

      \addplot[thick, magenta, densely dashed]
        table[x index=0, y index=2, col sep=space]
        {phase_shift_Im_4h_p_j13_pi-+_n26_jr5_nosrg_hw20_mv1.dat}
        node[pos=0.7, below, black, scale=0.8] {${}^2P^{+}_{1/2}$};

      \addplot[thick, green!50!black, only marks, mark repeat={10}]
        table[x index=0, y index=2, col sep=space]
        {phase_shift_4h_p_j13_pi-+_n28_jr5_nosrg_hw20_mv1.dat};

      \addplot[thick, magenta, only marks, mark repeat={10}, mark=triangle*]
        table[x index=0, y index=2, col sep=space]
        {phase_shift_Im_4h_p_j13_pi-+_n28_jr5_nosrg_hw20_mv1.dat};

    \nextgroupplot[
      title={$J=3/2$},
      ylabel style={yshift=-10pt},
      xlabel={$E_{\mathrm{c.m.}}$}
    ]

      \addplot[thick, green!50!black]
        table[x index=0, y index=3, col sep=space]
        {phase_shift_4h_p_j13_pi-+_n26_jr5_nosrg_hw20_mv1.dat}
        node[midway, below, black, scale=0.8] {${}^2D^{-}_{3/2}$};

      \addplot[thick, magenta, densely dashed]
        table[x index=0, y index=3, col sep=space]
        {phase_shift_Im_4h_p_j13_pi-+_n26_jr5_nosrg_hw20_mv1.dat}
        node[pos=0.7, above, black, scale=0.8] {${}^2D^{-}_{3/2}$};

      \addplot[thick, green!50!black, only marks, mark repeat={10}]
        table[x index=0, y index=3, col sep=space]
        {phase_shift_4h_p_j13_pi-+_n28_jr5_nosrg_hw20_mv1.dat};

      \addplot[thick, magenta, only marks, mark repeat={10}, mark=triangle*]
        table[x index=0, y index=3, col sep=space]
        {phase_shift_Im_4h_p_j13_pi-+_n28_jr5_nosrg_hw20_mv1.dat};

      \addplot[thick, green!50!black]
        table[x index=0, y index=4, col sep=space]
        {phase_shift_4h_p_j13_pi-+_n26_jr5_nosrg_hw20_mv1.dat}
        node[pos=0.5, below, black, scale=0.8] {${}^2P^{+}_{3/2}$};

      \addplot[thick, magenta, densely dashed]
        table[x index=0, y index=4, col sep=space]
        {phase_shift_Im_4h_p_j13_pi-+_n26_jr5_nosrg_hw20_mv1.dat}
        node[pos=0.45, above, black, scale=0.8] {${}^2P^{+}_{3/2}$};

      \addplot[thick, green!50!black, only marks, mark repeat={10}]
        table[x index=0, y index=4, col sep=space]
        {phase_shift_4h_p_j13_pi-+_n28_jr5_nosrg_hw20_mv1.dat};

      \addplot[thick, magenta, only marks, mark repeat={10}, mark=triangle*]
        table[x index=0, y index=4, col sep=space]
        {phase_shift_Im_4h_p_j13_pi-+_n28_jr5_nosrg_hw20_mv1.dat};

  \end{groupplot}
\end{tikzpicture}
\caption{Real (green) and imaginary (magenta) $\bar{p}$-${}^4 \rm He$ phase shifts calculated using the target's ground state. The regulator parameters $r_{\text{reg}}=7$ fm and $r_{\text{reg,c}}=5$ fm have been used.}
\label{fig:p_he4_phase_dep_nmax}
\end{figure}

In \autoref{fig:p_he4_phase_dep_nmax}, we show the real and imaginary parts of the $\bar{p}$-${}^4 \mathrm{He}$ scattering phase shifts for the $J=1/2$ (top) and $J=3/2$ (bottom) partial waves. The comparison between $N_{\text{max}}=26(27)$ and $28(29)$ demonstrates that our results are stable with respect to the model-space truncation. A small residual dependence on $r_{\text{reg}}$ remains, as discussed below for the scattering length, but it is under control. Reaching larger model spaces would further reduce this sensitivity, leaving the specific choice of cluster expansion as the main potential systematic bias, although this effect is expected to be limited for the more tightly bound ${}^4 \mathrm{He}$ target.
We note that partial waves with the same orbital angular momentum, such as ${}^{2}P^{+}_{1/2}$ and ${}^{2}P^{+}_{3/2}$, exhibit very similar phase shifts. This pattern is also reflected in other observables, including the level shifts and widths of the corresponding antiprotonic atom. \par
%
%
%
%
In the top panel of \autoref{tab:p_alpha_scat_dep}, we study the dependence of the $s$-wave scattering length on the model-space size. The convergence with respect to $N_{\text{max}}$ is satisfactory; however, the imaginary part still carries an estimated uncertainty below $3\%$, associated with the residual dependence on the regulator. We do not observe a clear reduction of this uncertainty when increasing $N_{\text{max}}$ from 26 to 30. Comparing this result with those obtained for lighter targets, we identify a general trend: The magnitude of the imaginary part of the $s$-wave scattering length decreases from the deuteron to ${}^4 \mathrm{He}$, whereas the real parts remain relatively similar for the $3 \leq A\leq5$ systems. \par
The $\alpha$ particle has no bound excited state. However, within the NCSM, one obtains discretized continuum states of the target, which we refer to as pseudostates. To assess their impact, we perform calculations with the target ground state only and with the ground state plus the first pseudostate (g.s.+$\alpha^{*}$). The corresponding results are shown in the bottom panel of \autoref{tab:p_alpha_scat_dep}. Since this additional channel is closed and lies far from the low-energy region used to extrapolate the scattering length, its contribution is found to be negligible. \par
%
%
\begin{table}
\caption{$\bar{p}$-${}^4 \rm He$ scattering length (in fm) for the ${}^2S_{1/2}^-$ channel. The top panel demonstrates the convergence with respect to $N_{\text{max}}$ using the target's ground state. The bottom panel investigates the dependence on the number of target states at $N_{\text{max}}=26$. Here g.s. refers to the calculation with the target ground state, while g.s.+$\alpha$* means the calculation with the ground state plus the first unbound excited state (pseudostate) in the same channel. The values in parentheses indicate the uncertainty due to dependence on the regulator parameter.}
\small
\centering
\begin{ruledtabular}
    \begin{tabular}{cc}
    $N_{\text{max}}=28$ & $N_{\text{max}}=30$ \\
    \midrule
    $1.62-0.70(\pm 0.02)i$ & $1.61-0.70(\pm 0.02)i$ \\
    \bottomrule \bottomrule
    g.s. & g.s.+$\alpha$* \\
    \midrule
    $1.61-0.70(\pm 0.01)i$ & $1.62-0.68(\pm 0.01)i$ \\
    \end{tabular}
\end{ruledtabular}
\label{tab:p_alpha_scat_dep}
\end{table}
%
%
\begin{table}
\caption{ $s$-wave (${}^2S_{1/2}^{-}$)  level shift and half-width (in keV) with $N_{\text{max}}=26$, calculated for the ground state ($n=1$) and the first excited state ($n=2$) of antiprotonic ${}^4 \rm He$. The $R$-matrix calculation is done with $a_c=200$ fm and $n_s=300$. For the Trueman results, we use the scattering lengths at the bottom panel of \autoref{tab:p_alpha_scat_dep} as input.}
\footnotesize
\centering
\begin{ruledtabular}
    \begin{tabular}{lcc}
    & $n=1$ & $n=2$ \\
    \midrule
    $R$-matrix (g.s.) &
      $21.7(\pm 0.1)-5.8(\pm 0.1)i$ &
      $3.03-0.92(\pm 0.02)i$ \\
    $R$-matrix (g.s.+$\alpha$*) &
      $21.7(\pm 0.1)-5.6(\pm 0.1)i$ &
      $3.03-0.89(\pm 0.01)i$ \\
    \midrule \midrule
    Trueman (g.s.) &
      $22.07-5.42i$ &
      $3.18-1.13i$ \\
    Trueman (g.s.+$\alpha$*) &
      $22.06-5.22i$ &
      $3.19-1.09i$ \\
    \end{tabular}
\end{ruledtabular}
\label{tab:p_he4_level}
\end{table}
%
%
In \autoref{tab:p_he4_level}, we show the $s$-wave level shift and half-width for the ground state ($n=1$) and the first excited state ($n=2$) of antiprotonic ${}^4 \rm He$. They can be obtained directly using the bound-state $R$-matrix method, and indirectly using the Deser~\cite{Deser} or Trueman formula~\cite{TRUEMAN196157}. For the indirect calculation, we have used the second-order Trueman relation~\cite{Carbonell:1992wd,carbonell2023comparison}, which uses the Coulomb-modified $\bar{p}$-${}^4 \rm He$ scattering length as input. Similar to the scattering length, the effect of including the first pseudostate (g.s.+$\alpha$*) is restricted to the imaginary part, and less than $4\%$. At this value of $N_{\text{max}}$, the $R$-matrix results have an uncertainty due to regulators, estimated to be less than $3\%$. The difference between the $R$-matrix and Trueman results comes from the large expansion parameter used in the Trueman relation. In other words, as the atomic radius decreases, the second-order Trueman relation loses its validity, and one needs to go to higher orders for the expansion to converge. Similarly, we report the convergence check for the 2P ($n=2,\, \ell=1$) level shifts and half-widths in \autoref{tab:level_he4_pwave}. Our results for the ${}^2P_{1/2}^{+}$ and ${}^2P_{3/2}^{+}$ channels are nearly identical.
%
%
\begin{table}
\caption{2P ($n=2,\, \ell=1$) atomic level shifts and half-widths (in eV) for antiprotonic ${}^4 \rm He$ obtained through bound-state $R$-matrix calculation at different values of $N_{\text{max}}$. We use $a_c=400$ fm and $n_s=300$.}
\small
\centering
\begin{ruledtabular}
  \begin{tabular}{ccc}
    \multicolumn{1}{c}{Channel} &
    \multicolumn{1}{c}{$N_{\text{max}}=28$} &
    \multicolumn{1}{c}{$N_{\text{max}}=30$} \\
    \midrule
    ${}^2P_{1/2}^+$ &
      $8.5(\pm 0.2)-21(\pm 2)i$ &
      $8.5(\pm 0.2)-21(\pm 1)i$ \\
    ${}^2P_{3/2}^+$ &
      $8.4(\pm 0.2)-21(\pm 2)i$ &
      $8.4(\pm 0.2)-21(\pm 1)i$ \\
  \end{tabular}
\end{ruledtabular}
\label{tab:level_he4_pwave}
\end{table}
%
%

\end{document}